\documentclass[aps,prd,twocolumn,showpacs,floatfix]{revtex4}
\usepackage{bm}
\usepackage{graphicx}
\newcommand{\Tr}{\mbox{Tr}}
\newcommand{\ReTr}{\mbox{ReTr}}
\begin{document}
\title{Analytic Smearing of $\bm{SU(3)}$ Link Variables in Lattice QCD}
\author{Colin Morningstar}
\affiliation{Department of Physics, Carnegie Mellon University,
             Pittsburgh, PA 15213, USA}
\author{Mike Peardon}
\affiliation{School of Mathematics, Trinity College, Dublin 2, Ireland}
\date{\today}

\begin{abstract}
An analytic method of smearing link variables in lattice QCD is proposed and
tested.  The differentiability of the smearing scheme with respect to the
link variables permits the use of modern Monte Carlo updating methods based
on molecular dynamics evolution for gauge-field actions constructed using
such smeared links.  In examining the smeared mean plaquette and the static
quark-antiquark potential, no degradation in effectiveness is observed as
compared to link smearing methods currently in use, although an increased
sensitivity to the smearing parameter is found.
\end{abstract}
\maketitle

\section{Introduction}
The extraction of hadron masses and matrix elements from
Monte Carlo estimates of Euclidean-space correlation functions
in lattice QCD can be done more reliably and accurately when
operators which couple more strongly to the states of interest
and less strongly to the higher-lying contaminating states
are used.  For states containing gluons, a crucial ingredient
in constructing such operators in lattice QCD is link
variable smearing.  Operators constructed out of
smeared or fuzzed links have dramatically reduced mixings with
the high frequency modes of the theory.  The use of such operators
has been shown to especially benefit determinations of the
glueball spectrum\cite{colin2}, hybrid meson masses\cite{hybrids,jkm1},
the torelon spectrum\cite{torelons}, and excitations of the static
quark-antiquark potential\cite{jkm2}.

Link variable smoothing is also playing an increasingly important
role in the construction of improved lattice actions.  In Ref.~\cite{fat1},
the use of so-called fat links in a staggered quark action was shown
to significantly decrease flavor symmetry breaking.
Smeared links were subsequently used\cite{fat2} to construct hypercubic
fermion actions having improved rotational invariance.
A staggered fermion action using a link smearing transformation\cite{hyp}
known as hypercubic (HYP) fat links was shown to improve flavor symmetry
by an order of magnitude relative to the standard action.
The so-called Asqtad improved staggered 
quark action\cite{asqtad1,asqtad2,asqtad3}
also makes use of link fattening to reduce flavor symmetry breaking.
Another variant of fermion actions which exploit smeared link variables
is the fat link irrelevant clover (FLIC) action\cite{flic}.
FLIC fermions are Wilson-like and described by an action which
includes an irrelevant clover improvement term constructed using
smeared links.  Fat links have also been used to construct a
gauge action\cite{fatgauge} with reduced discretization errors using
approximate renormalization group transformations.

The link fuzzing algorithm most often used in gluonic operator construction
is that described in
Ref.~\cite{APEsmear} in which every spatial link $U_j(x)$ on the lattice
is replaced by itself plus a real weight $\rho$ 
times the sum of its four neighboring
(spatial) staples, projected back into $SU(3)$.  Such a fuzzing step is
iterated $n_\rho$ times to obtain the final fuzzed link variables.
Empirically, one finds that the projection into $SU(3)$ is a crucial
ingredient of the smearing.  Link smearings which do not apply such
a projection are found to be much less effective.  The projection into
$SU(3)$ is not unique and must be carefully defined so that all symmetry
properties of the link variables are preserved.  Various ways of implementing
this projection have appeared in the literature.  Given a $3\times 3$
matrix $V$, its projection $U$ into $SU(3)$ is often taken to be the
matrix $U\in SU(3)$ which maximizes $\ReTr(UV^\dagger)$.  An iterative
procedure is required to perform such a maximization.  
Alternatively\cite{su3projectA}, the projected matrix $U$ can be defined by
$ U = V\ (V^\dagger V)^{-1/2} \det(V^{-1}V^\dagger)^{1/6}.$ 

Although the above smearing procedure works well in practice, the
projection is somewhat unpalatable since it may be viewed as an
arbitrary and abrupt way to remain within the group.  More importantly,
the branch cuts and lack of differentiability inherent in the projection
can hinder or even make impossible the application of modern Monte Carlo
updating techniques, such as hybrid Monte Carlo (HMC)\cite{hmc}, which
require knowing the response of the action to a small change in one of
the link variables.

A link smearing method which circumvents these problems is proposed
and tested in this paper.  The link smearing method is analytic
everywhere in the finite complex plane and utilizes the exponential
function in such a manner to remain within $SU(3)$, eliminating the
need for any projection back into the group.  Because of this construction,
the algorithm is useful for any Lie group.  The method is described
in Sec.~\ref{sec:analytic} and its practical implementations when
constructing operators and when computing the response of the action
to a change in a link variable are detailed in 
Secs.~\ref{sec:implement} and \ref{sec:force}, respectively.
Numerical tests of the smearing are presented in Sec.~\ref{sec:tests}.
Using the smeared plaquette and the effective mass $aE_{\rm eff}(t)$
associated with the static quark-antiquark potential, no
degradation in effectiveness is observed as compared to the standard
link smearing method, although an increased sensitivity to the
smearing parameter is found.  The results also suggest that lattice
actions and operators constructed out of smeared link variables may
be much less afflicted by radiative corrections since the
usually-dominant large tadpole contributions are drastically reduced.

\section{Analytic link smearing}
\label{sec:analytic}

A method of smearing link variables which is analytic, and hence
differentiable, everywhere in the finite complex plane
can be defined as follows.
Let $C_\mu(x)$ denote the weighted sum of the perpendicular staples which
begin at lattice site $x$ and terminate at neighboring site 
$x\!+\!\hat{\mu}$:
\begin{eqnarray}
 C_\mu(x)&=&\sum_{\nu\neq \mu}\rho_{\mu\nu}\biggl(
 U_\nu(x) U_\mu(x\!+\!\hat{\nu}) U_\nu^\dagger(x\!+\!\hat{\mu})\nonumber\\
&&+ U^\dagger_\nu(x\!-\!\hat{\nu}) U_\mu(x\!-\!\hat{\nu})
  U_\nu(x\!-\!\hat{\nu}\!+\!\hat{\mu})
\biggr), \label{eq:Cdef}
\end{eqnarray}
where $\hat{\mu},\hat{\nu}$ are vectors in directions $\mu,\nu$,
respectively, having the length
of one lattice spacing in that direction (fundamental lattice
translation vectors).  The weights $\rho_{\mu\nu}$ are tunable real
parameters. Then the matrix $Q_\mu(x)$, defined in $SU(N)$ by
\begin{eqnarray}
Q_\mu(x) &=& 
  \frac{i}{2}\Bigl(\!\Omega^\dagger_\mu(x)\!-\!\Omega_\mu(x)\!\Bigr)
   \!-\!\frac{i}{2N}\Tr\Bigl(\!\Omega^\dagger_\mu(x)
  \!-\!\Omega_\mu(x)\!\Bigr),\nonumber\\
\Omega_\mu(x) &=& C_\mu(x)\ U_\mu^\dagger(x),
   \quad\mbox{(no summation over $\mu$)} \label{eq:Qdef}
\end{eqnarray}
is Hermitian and traceless, and hence, $e^{i Q_\mu(x)}$ is an
element of $SU(N)$.  
We use this fact to define an iterative, analytic link
smearing algorithm in which the links $U_\mu^{(n)}(x)$ at step $n$
are mapped into links $U_\mu^{(n+1)}(x)$ using
\begin{equation}
U^{(n+1)}_\mu(x)=\exp\Bigl(i Q_\mu^{(n)}(x)\Bigr)\ U_\mu^{(n)}(x).
\label{eq:Ustout}
\end{equation}
The fact that $e^{iQ_\mu(x)}$
is an element of $SU(N)$ guarantees that $U^{(n+1)}_\mu(x)$ is also an element
of $SU(N)$, eliminating the need for a projection back onto the gauge group.
This fuzzing step can be iterated $n_\rho$ times to finally
produce link variables which we call {\em stout links}\cite{pub},
denoted by $\tilde{U}_\mu(x)$:
\begin{equation}
 U\rightarrow U^{(1)}\rightarrow U^{(2)}\rightarrow\cdots\rightarrow
  U^{(n_\rho)}\equiv \tilde{U}.
\end{equation}

One common choice of the staple weights is
\begin{equation}
   \rho_{jk}=\rho,\qquad \rho_{4\mu}=\rho_{\mu 4}=0,
\end{equation}
which yields a three-dimensional scheme in which only the spatial
links are smeared.  Another common choice is an isotropic four-dimensional
scheme in which all weights are chosen to be the same
$\rho_{\mu\nu}=\rho$.  Note that such a scheme 
was used in Ref.~\cite{luscher} as a field transformation
to eliminate a certain interaction term from the most general $O(a^2)$
on-shell improved gauge action.

It is not difficult to show that, given an appropriate choice of weights
$\rho_{\mu\nu}$, the stout links have symmetry transformation
properties identical to those of the original link variables.
Under any local gauge transformation $G(x)$,
the link variables transform according to 
$U_\mu(x)\rightarrow G(x)\ U_\mu(x)\ G^\dagger(x\!+\!\hat{\mu})$,
where the $G(x)$ are $SU(N)$ matrices.  Thus,
under such a gauge transformation,
\begin{eqnarray}
Q_\mu(x)&\rightarrow& G(x)\ Q_\mu(x)\ G^\dagger(x),\\
e^{iQ_\mu(x)}&\rightarrow& G(x)\ e^{iQ_\mu(x)}\ G^\dagger(x),\\
U^{(1)}_\mu(x)&\rightarrow& G(x)\ e^{iQ_\mu(x)} G^\dagger(x)
 G(x)U_\mu(x)\ G^\dagger(x\!+\!\hat{\mu})\nonumber\\
 &=& G(x)\ U^{(1)}_\mu(x)\ G^\dagger(x\!+\!\hat{\mu}),
\end{eqnarray}
as required.  One can also easily show that if the weights $\rho_{\mu\nu}$
respect the rotation and reflection symmetries of the lattice,
the stout links obey the
required transformation properties under all rotations and all
reflections in a plane containing the link.  Lastly, consider a
reflection in a plane normal to a link $U_\mu(x)$ and passing through its
midpoint $x\!+\!\frac{1}{2}\hat{\mu}$. Under such an operation, the
link $U_\mu(x)$ transforms according to 
$U_\mu(x)\rightarrow U_{-\mu}(x\!+\!\hat{\mu})=U_\mu^\dagger(x)$ and
for $\nu\neq \mu$, $U_\nu(x)\leftrightarrow U_\nu(x\!+\!\hat{\mu})$ and
$U_\nu^\dagger(x\!+\!\hat{\mu})\leftrightarrow U_\nu^\dagger(x)$, yielding
$C_\mu(x)\rightarrow C_\mu^\dagger(x)$ assuming the $\rho_{\mu\nu}$
are real, so that
\begin{eqnarray}
 \Omega_\mu(x) &\rightarrow & C_\mu^\dagger(x)U_\mu(x)=U_\mu^\dagger(x)
 \Omega_\mu^\dagger(x) U_\mu(x),\\
 Q_\mu(x)&\rightarrow & -U_\mu^\dagger(x)Q_\mu(x)U_\mu(x),\\
 U^{(1)}_\mu(x)&\rightarrow & U_\mu^\dagger(x) e^{-iQ_\mu(x)}
   = U_\mu^{(1)\dagger}(x),
\end{eqnarray}
as required.  Given that $U^{(1)}_\mu(x)$ transforms under all symmetry
operations in the same manner as $U_\mu(x)$, it follows that the stout
links $\tilde{U}_\mu(x)$ have symmetry transformation properties identical
to those of the original link variables.

\begin{figure}
\includegraphics[width=3.35in,bb=0 0 598 199]{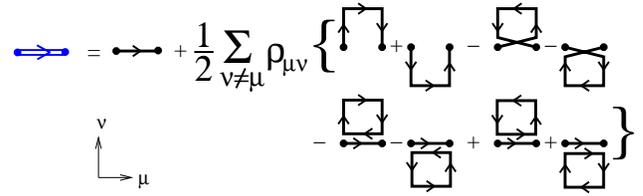}
\caption[figF]{The expansion up to first order in the $\rho_{\mu\nu}$
 of the new link variable
 $U^{(1)}$ in terms of paths of the original links.  Each closed loop
 includes a trace with a factor $1/N$ in $SU(N)$.
\label{fig:paths}}
\end{figure}

Since the exponential function has a power series expansion with
an infinite radius of convergence, each stout link may be viewed as
an incredibly large and complicated sum of paths.  For small $\rho_{\mu\nu}$,
the paths which make up the link variable $U^{(1)}$ to first order
in the $\rho_{\mu\nu}$ are shown in Fig.~\ref{fig:paths}.  Note that the 
standard smearing method, defined by
\begin{equation} 
U^{(1)}_\mu(x) = {\cal P}_{SU(3)}\ \biggl\{ 
 U_\mu(x) + C_\mu(x) \biggr\},
\label{eq:Ufuzz}
\end{equation}
where ${\cal P}_{SU(3)}$ denotes the projection into $SU(3)$,
yields the same sum of paths at first order in $\rho_{\mu\nu}$.

A few further remarks are worthy of note.  First, an alternative smearing
scheme in which the weights $\rho_{\mu\nu}$ are chosen to be imaginary
does not reduce the couplings to the high-frequency modes of the theory.
Secondly, it is not possible to remove the leading $O(a^2)$ discretization
errors in the Wilson gauge action by expressing the action in terms
of stout links; diminishing the $O(\alpha_s a^2)$ errors by reducing their
tadpole contributions is the best which can be achieved.  Thirdly, the
use of stout links in Symanzik-improved actions may eliminate the need
to use tadpole-improvement in determining the couplings in the action.
Lastly, the definition of $C_\mu(x)$ given in Eq.~(\ref{eq:Cdef}) is not
unique. Analytic smearing schemes can be constructed as outlined above
using sums of other paths from site $x$ to $x\!+\!\hat{\mu}$, such as
those used in fat link\cite{fat1} and HYP\cite{hyp} smearings.

\section{Implementation of the smearing}
\label{sec:implement}

To implement this analytic link smearing scheme, the efficient evaluation
of $\exp( i Q)$, where $Q$ is a traceless Hermitian $3\times 3$ matrix,
is required. The Cayley-Hamilton theorem states that every matrix is a zero
of its characteristic polynomial, so that
\begin{equation}
  Q^3 - c_1\ Q - c_0\ I = 0,
\label{eq:CayHam}
\end{equation}
where
\begin{eqnarray}
 c_0 &=&  \det Q=\textstyle\frac{1}{3}\Tr(Q^3),\\
 c_1 &=& \textstyle\frac{1}{2} \Tr(Q^2)\geq 0.
\end{eqnarray}
The Hermiticity of $Q$ requires $27c_0^2-4c_1^3\leq 0$  and
the definition of $Q$ given in Eq.~(\ref{eq:Qdef}) restricts the possible
values of $c_1$.  Thus, the coefficients $c_0$ and $c_1$ satisfy
\begin{equation}
 -c_0^{\rm max}\leq c_0 \leq c_0^{\rm max},\qquad
0 \leq c_1 \leq c_1^{\rm max},
\end{equation}
where
\begin{eqnarray}
 c_0^{\rm max}&=&2\left(\frac{c_1}{3}\right)^{3/2},\\
  c_1^{\rm max} &=& \frac{1}{32}
  (69+11\sqrt{33})\Bigl(\displaystyle\sum_{\nu\neq\mu}
\rho_{\mu\nu}\Bigr)^2,
\end{eqnarray}
for each $\mu$.  Eq.~(\ref{eq:CayHam}) implies that $Q^n$ for integer
$n\geq 3$ can be expressed in terms of $Q^2$, $Q$, and the identity
matrix $I$.  Hence, we can write
\begin{equation}
 e^{i Q} = f_0\ I + f_1\ Q + f_2\ Q^2,
\label{eq:fexpand}
\end{equation}
where the three scalar coefficients $f_j=f_j(c_0,c_1)$ are
basis independent, depending only on $c_0$ and $c_1$.  Eq.~(\ref{eq:fexpand})
is valid for {\em any} $3\times 3$ Hermitian, traceless matrix $Q$.

Let $q_1,q_2,q_3$ denote the three eigenvalues of $Q$. Since $Q$ is
Hermitian and traceless, we know that these are real numbers satisfying
$q_1+q_2+q_3=0$.  These eigenvalues are the three roots of a cubic
polynomial which can be easily determined:
\begin{eqnarray}
q_1 &=& 2u,\\
q_2 &=& -u+w,\\
q_3 &=& -q_1-q_2=-u-w,
\end{eqnarray}
where
\begin{eqnarray}
u &=& \sqrt{\textstyle\frac{1}{3}c_1}
  \ \cos\left(\textstyle\frac{1}{3}\theta\right),\label{eq:udef}\\
w &=& \sqrt{c_1}\ \sin\left(\textstyle\frac{1}{3}\theta\right),\label{eq:wdef}\\
\theta &=& \arccos\left(\frac{c_0}{c_0^{\rm max}}\right).\label{eq:thetadef}
\end{eqnarray}
Given these eigenvalues, the matrix $Q$ can be written
\begin{equation}
 Q = M\ \Lambda_Q\ M^{-1},\qquad 
\Lambda_Q= \left[\begin{array}{ccc} q_1 &0 & 0\\ 0 & q_2 & 0\\
 0 & 0 & q_3\end{array}\right] ,
\end{equation}
where $M$ is a unitary matrix.  Then it easily follows that
\begin{equation}
  e^{i \Lambda_Q}=f_0 I + f_1\Lambda_Q + f_2\Lambda_Q^2.
\end{equation}
Explicitly, we have the following linear system of equations to solve:
\begin{equation}
 \left[\begin{array}{ccc} 1 & q_1 & q_1^2 \\
  1 & q_2 & q_2^2 \\  1 & q_3 & q_3^2 
 \end{array}\right]\left[\begin{array}{c}
  f_0\\f_1\\f_2\end{array}\right]=\left[\begin{array}{c}
 e^{i q_1}\\e^{i q_2}\\e^{i q_3}\end{array}\right].
\label{eq:system}\end{equation}
If all three eigenvalues are distinct, this system of equations
has a unique solution, but when two of the eigenvalues are exactly the
same, the solution to Eq.~(\ref{eq:system}) is not unique and one of the
$f_j$'s can be freely set to any value. The case of degenerate eigenvalues
occurs when $c_0\rightarrow \pm c_0^{\rm max}$.  

In practice, it is extremely unlikely that an exact degeneracy will be
encountered during a numerical simulation.  Much more likely is the
possibility that two eigenvalues are nearly, but not quite, equal.
Encounters with both near and exact degeneracies can be handled
by expressing the $f_j$ in terms of $u$
and $w$ in order to isolate and tame the numerically sensitive part.
One finds that the $f_j$ factors can be written
\begin{equation}
   f_j = \frac{h_j}{(9u^2-w^2)},
\label{eq:fdef1}
\end{equation}
where the $h_j$ are well-behaved functions given by
\begin{eqnarray}                                           
  h_0 &=& (u^2\!-\!w^2)e^{2i u}+e^{-i u}\bigl\{ 8u^2\cos( w)\nonumber\\
  &&\quad + 2iu(3u^2\!+\!w^2)\xi_0( w)\bigr\}, \\
  h_1 &=& 2ue^{2i u}-e^{-i u}\bigl\{ 2u\cos( w) \nonumber\\
  &&\quad - i(3u^2\!-\!w^2)\xi_0( w)\bigr\}, \\
  h_2 &=& e^{2i u}-e^{-i u}\left\{ \cos( w)
  + 3iu\xi_0( w)\right\},
\label{eq:hdef}
\end{eqnarray}
defining
\begin{equation}
 \xi_0(w) = \frac{\sin w}{w}.
\end{equation}
The $w\rightarrow 0$ problem as $c_0\rightarrow c_0^{\rm max}$
has been completely contained in $\xi_0$.  To numerically evaluate 
this factor, one uses, for example,
\[
 \xi_0(w) = \left\{ \begin{array}{ll}
  1\!-\!\frac{1}{6}w^2\left(1\!-\!\frac{1}{20}w^2
  \left(1\!-\!\frac{1}{42}w^2\right)\right) ,
 & \  \vert w\vert\leq 0.05,\\[2mm]
  \sin(w)/w, & \ \vert w\vert>0.05.
 \end{array}\right.
\]
However, the $w\rightarrow 3u\rightarrow \sqrt{3}/2$ limit as
$c_0\rightarrow -c_0^{\rm max}$ has not been tamed.
Fortunately, this problem can be circumvented using the following
symmetry relations under $c_0\rightarrow -c_0$:
\begin{equation}
   f_j(-c_0,c_1)=(-)^j\ f_j^\ast(c_0,c_1). \label{eq:fsym}
\end{equation}
Thus, the determination of the $f_j$ coefficients for $c_0<0$ can
be achieved by computing the coefficients for $\vert c_0\vert$
and utilizing Eq.~(\ref{eq:fsym}).  This means that a
situation in which the denominator in Eq.~(\ref{eq:fdef1})
becomes nearly zero will never be encountered.  It should be
emphasized that the $f_j$ coefficients are smooth, non-singular
functions of $c_0$ and $c_1$.  There are no actual singularities
as $c_0\rightarrow\pm c_0^{\rm max}$.  The numerical evaluation
of the $f_j$ coefficients to machine precision in these limits
simply requires additional care.

As an aside, if one writes $Q=\frac{1}{2}\sum_{a=1}^8 Q_a\lambda^a$
in terms of the eight Gell-Mann matrices $\lambda^a$, then
\begin{eqnarray}
  e^{i Q}&=&u_0+\textstyle\frac{1}{2}\sum_{a=1}^8 u_a \lambda^a,\\
  u_0 &=& f_0+\textstyle\frac{2}{3}c_1f_2,\\
  u_a &=& f_1 Q_a + \textstyle\frac{1}{2} f_2 d^{abc}Q_bQ_c,\\
  c_0 &=& \textstyle\frac{1}{12}d^{abc}Q_a Q_bQ_c,\\
  c_1 &=& \textstyle\frac{1}{4}Q_aQ_a,
\end{eqnarray}
where $d^{abc}$ and $f^{abc}$ are the real symmetric and antisymmetric
structure constants, respectively.

\section{Molecular dynamics evolution}
\label{sec:force}

The analyticity of the stout-link scheme permits the use of modern Monte
Carlo updating methods for gauge-field actions constructed using stout links.
Molecular dynamics evolution forms the core of many of the Markov processes
used to generate ensembles of field configurations needed for unquenched
QCD simulations\cite{hmcgauge,hmcfermion,hmc,Ralgorithm}.
Since a small change to the underlying gauge
fields always leads to a small change in the stout links, then provided
the small change to the stout links causes small changes to the action of
the theory, the force term on the underlying links is always well-defined. 
The importance of defining a link smearing scheme which permits the
computation of the molecular dynamics force term has been emphasized
recently in Ref.~\cite{flic2}.

The $\Sigma_\mu(x)$ force field describes how the action $S$ changes
when the gauge-field link variables $U_\mu(x)$ change, holding the momenta 
$P_\mu(x)$ and the pseudofermion field $\phi(x)$ fixed.
For this reason, we shall write the action $S=S[U]$ in this section,
ignoring its dependence on the pseudofermion field.  The transpose
of the force field is defined by the derivative of the action with
respect to the link variables:
\begin{equation}
 \Sigma_\mu(x)=\left(\frac{\partial S[U]}{\partial U_\mu(x)}\right)^T.
\end{equation}
Assume that the action can be written as a sum of
a term $S_{\rm th}[U]$ constructed entirely of the original thin
link variables and a term $S_{\rm st}[\tilde{U}]$ constructed
completely out of stout links:
\begin{equation}
  S[U]=S_{\rm th}[U]+S_{\rm st}[\tilde{U}].
\end{equation}
Then the force field may be written
\begin{equation}
  \Sigma_\mu(x)=\Sigma_\mu^{\rm(th)}(x)+\Sigma_\mu^{(0)}(x),
\end{equation}
where
\begin{equation}
 \Sigma_\mu^{\rm(th)}(x)=\left(\frac{\partial S_{\rm th}[U]}{
  \partial U_\mu(x)}\right)^T\!\!\!, \quad
 \Sigma_\mu^{\rm(0)}(x)=\left(\frac{\partial S_{\rm st}[\tilde{U}]}{
  \partial U_\mu(x)}\right)^T\!\!\!.
\end{equation}
The computation of $\Sigma_\mu^{\rm(th)}(x)$ is usually
straightforward, and nothing more about its determination needs
discussion here.  We now focus on the evaluation of 
$\Sigma_\mu^{\rm(0)}(x)$. 

Recall that the stout links are constructed iteratively starting
with the original links 
$U\rightarrow U^{(1)}\rightarrow U^{(2)}
 \rightarrow\cdots\rightarrow U^{(n_\rho)}=\tilde{U}$.  
The computation of the force field proceeds similarly in an
iterative fashion, except that the order is reversed
$\tilde\Sigma=\Sigma^{(n_\rho)}\rightarrow\Sigma^{(n_\rho\!-\!1)}
\rightarrow\cdots\rightarrow\Sigma^{(1)}\rightarrow\Sigma^{(0)}$.
The sequence starts by computing
\begin{equation}
 \tilde{\Sigma}_\mu(x)=\left(\frac{\partial S_{\rm st}[\tilde{U}]}{
  \partial \tilde{U}_\mu(x)}\right)^T.
\end{equation}
This step depends on the form of the gauge and fermion action in terms
of the stout links and is usually just as straightforward as the
computation of $\Sigma_\mu^{\rm(th)}(x)$.  In subsequent steps,
$\Sigma^{(k)}$ and $U^{(k\!-\!1)}$ are used to compute
$\Sigma^{(k\!-\!1)}$, where the effective force at level $k$ is
defined by
\begin{equation}
  \Sigma_\mu^{(k)}(x)=\left(\frac{\partial S_{\rm st}[\tilde{U}]}{
  \partial U_\mu^{(k)}(x)}\right)^T.
\end{equation}
The recursive mapping
\begin{equation}
   \left\{  \Sigma^{(k)}, U^{(k-1)}\right\} 
  \longrightarrow \Sigma^{(k-1)}
 \label{eq:mapping}
\end{equation}
is repeated until $\Sigma^{(0)}$ is finally evaluated.

In Monte Carlo methods based on molecular dynamics, such as HMC\cite{hmc}
and the $R$-algorithm\cite{Ralgorithm}, the flow of lattice configurations
through phase space via Hamilton's equations is 
parametrized by a fictitious simulation time
coordinate $\tau$.  To determine the mapping in Eq.~(\ref{eq:mapping}),
it is convenient to express the force field in terms of the rate of change
of the action with respect to this time coordinate $\tau$:
\begin{equation}
  \frac{d}{d\tau}S_{\rm st}[\tilde{U}] = 2\sum_{x,\mu} \ReTr \left\{ 
        \Sigma_\mu^{(k)}(x)   \;
        \frac{d}{d\tau} U^{(k)}_\mu(x) 
\right\}, 
\end{equation}
for $k=0,1,\dots,n_\rho$. One then proceeds using the chain rule of
differentiation:
\begin{eqnarray}
\frac{dU^{(k)}_\mu(x)}{d\tau} &=&  e^{i Q^{(k-1)}_\mu(x)} 
   \frac{dU^{(k-1)}_\mu(x)}{d\tau} \nonumber\\
   &&+ \frac{d(e^{i Q^{(k-1)}_\mu(x)})}{d\tau} U^{(k-1)}_\mu(x).
\end{eqnarray}
To simplify matters, $Q^{(k-1)}_\mu(x),\ U^{(k-1)}_\mu(x),\ \Sigma^{(k-1)}_\mu(x)$
shall simply be written as $Q,U,\Sigma$, respectively,
in the calculations which follow, and $\Sigma^{(k)}_\mu(x)$ and
$U^{(k)}_\mu(x)$ shall be written as $\Sigma^\prime$ and $U^\prime$,
respectively.  The Cayley-Hamilton theorem
gives us
\begin{eqnarray}
    \frac{d(e^{i Q})}{d\tau} &=& \frac{d}{d\tau}
             \left(f_0 + f_1 Q + f_2 Q^2\right),\nonumber\\
   &=& \frac{df_0}{d\tau}+\frac{df_1}{d\tau}Q+\frac{df_2}{d\tau}Q^2\nonumber\\
    &&   + f_1\frac{dQ}{d\tau}+f_2\frac{dQ}{d\tau}Q+f_2Q\frac{dQ}{d\tau}.
\end{eqnarray}
Since the $f_j$ coefficients are functions $f_j=f_j(u,w)$ of $u$
and $w$ only, one has
\begin{equation}
    \frac{df_j}{d\tau} = \left(\frac{\partial f_j}{\partial u}\right)
   \frac{du}{d\tau}
    + \left(\frac{\partial f_j}{\partial w}\right)
   \frac{dw}{d\tau},
\end{equation}
and since $u$ and $w$ are functions of $c_0$ and $c_1$ only, then
\begin{eqnarray}
 \frac{du}{d\tau} &=& \left(\frac{\partial u}{\partial c_0}\right)
   \frac{dc_0}{d\tau}
                     +\left(\frac{\partial u}{\partial c_1}\right)
   \frac{dc_1}{d\tau},\\
 \frac{dw}{d\tau} &=& \left(\frac{\partial w}{\partial c_0}\right)
    \frac{dc_0}{d\tau}
                     +\left(\frac{\partial w}{\partial c_1}\right)
   \frac{dc_1}{d\tau}.
\end{eqnarray}
Next, one finds that
\begin{eqnarray}
    \frac{dc_0}{d\tau} &=& {\textstyle\frac{1}{3}}
    \frac{d}{d\tau} \Tr(Q^3) = 
 \Tr\left(Q^2 \frac{dQ}{d\tau}\right),\\
    \frac{dc_1}{d\tau} &=& {\textstyle\frac{1}{2}}
   \frac{d}{d\tau} \Tr (Q^2) = 
\Tr\left(Q \frac{dQ}{d\tau}\right),
\end{eqnarray}
and 
\begin{equation}
\begin{array}{ll}
\displaystyle\frac{\partial u}{\partial c_0}=\frac{1}{2(9u^2-w^2)}, &
\ \displaystyle \frac{\partial u}{\partial c_1}=\frac{u}{(9u^2-w^2)}, \\[5mm]
\displaystyle\frac{\partial w}{\partial c_0}=\frac{-3u}{2w(9u^2-w^2)}, &
\ \displaystyle \frac{\partial w}{\partial c_1}=\frac{3u^2-w^2}{2w(9u^2-w^2)}.
\end{array}
\end{equation}
Now define
\begin{equation}
r^{(1)}_j = \frac{\partial h_j}{\partial u},\qquad
r^{(2)}_j =\frac{1}{w}\frac{\partial h_j}{\partial w},
\end{equation}
and
\begin{eqnarray}
 b_{1j} &=& \frac{2ur_j^{(1)}+
    (3u^2\!-\!w^2)r^{(2)}_j-2(15u^2\!+\!w^2)f_j}{2(9u^2-w^2)^2},\\
 b_{2j}&=& \frac{r_j^{(1)}-3ur^{(2)}_j-24uf_j}{
   2(9u^2-w^2)^2},
\end{eqnarray}
then
\begin{equation}
 \frac{df_j}{d\tau}=b_{1j}\Tr\left(Q\frac{dQ}{d\tau}\right)
   +b_{2j}\Tr\left(Q^2\frac{dQ}{d\tau}\right),
\end{equation}
where
\begin{eqnarray}
r^{(1)}_0 &=& 2\Bigl(u+i(u^2\!-\!w^2)\Bigr)e^{2i u}\nonumber\\
  &&+2e^{-i u}\Bigl\{
 4u\bigl(2\!-\!i u  \bigr)\cos w \nonumber\\
  && +i\bigl(9u^2+w^2-i u( 3u^2+w^2) \bigr)\xi_0( w) 
  \Bigr\}, \\
r^{(1)}_1 &=& 2(1+2i u)e^{2i u}+e^{-i u}\Bigl\{
    -2(1-i u) \cos w \nonumber\\
  &&  +i\bigl( 6u +i(w^2-3u^2) \bigr)\xi_0( w) 
   \Bigr\}, \\
r^{(1)}_2 &=&  2i e^{2i u}+i e^{-i u}\Bigl\{
   \cos w -3 (1-i  u)  \xi_0( w)
   \Bigr\},\\
r^{(2)}_0 &=& -2e^{2i u}+2i ue^{-i u}\Bigl\{
  \cos w \nonumber\\
  && +( 1+4i u) \xi_0( w) 
  +3u^2\xi_1( w)\Bigr\}, \\
r^{(2)}_1 &=& -i e^{-i u}\Bigl\{
   \cos w  +(1+2i u)\xi_0( w)\nonumber\\ 
  &&-3u^2\xi_1( w)\Bigr\}, \\
r^{(2)}_2 &=&   e^{-i u}\Bigl\{
     \xi_0( w) 
   -3i  u \xi_1( w)\Bigr\},
\end{eqnarray}
with
\begin{eqnarray}
 \xi_0(w) &=& \frac{\sin w}{w},\\
 \xi_1(w) &=& \frac{\cos w}{w^2}-\frac{\sin w}{w^3}.
\end{eqnarray}
Given the above results,
\begin{eqnarray}
\frac{d(e^{i Q})}{d\tau}&=&\Tr\left(Q\frac{dQ}{d\tau}\right) B_1
+\Tr\left(Q^2\frac{dQ}{d\tau}\right) B_2\nonumber\\
&&+ f_1\frac{dQ}{d\tau}+f_2\frac{dQ}{d\tau}Q+f_2Q\frac{dQ}{d\tau},
\label{eq:dRdef}
\end{eqnarray}
where
\begin{equation}
 B_i = b_{i0}+b_{i1}\ Q + b_{i2}\ Q^2.
\end{equation}
The numerically sensitive $w\rightarrow 0$ limit has been
totally absorbed into $\xi_0$ and $\xi_1$, and
numerical problems as $c_0\rightarrow -c_0^{\rm max}$ can 
be circumvented by exploiting the following symmetry relation:
\begin{equation}
  b_{ij}(-c_0,c_1)=
  (-)^{i+j+1}\ b_{ij}^\ast(c_0,c_1). \label{eq:bsym}
\end{equation}
Once again, we emphasize that these coefficients are well-behaved,
non-singular functions of $c_0$ and $c_1$.

The rate of change of the stout links with respect to the simulation
time is then given by
\begin{eqnarray}
  \frac{dU^\prime}{d\tau} &=& e^{i Q} \frac{dU}{d\tau} 
  + \biggl\{  \Tr\left(Q\frac{dQ}{d\tau}\right) B_1
   +\Tr\left(Q^2\frac{dQ}{d\tau}\right) B_2 \nonumber\\
   &&\qquad + f_1\frac{dQ}{d\tau}+f_2\frac{dQ}{d\tau}Q+f_2Q\frac{dQ}{d\tau}
 \biggr\} U,
\end{eqnarray}
and hence, 
\begin{equation}
\ReTr\left(\Sigma^\prime
  \frac{dU^\prime}{d\tau}\right)
 =\ReTr\left(\Sigma^\prime e^{i Q} \frac{dU}{d\tau}\right)
                    - \ReTr\left(i\Lambda \frac{d\Omega}{d\tau}\right),
   \label{eqn:diff-stouttrace}
\end{equation}
defining 
\begin{eqnarray}
   \Lambda &=& 
   \frac{1}{2} (\Gamma + \Gamma^\dagger) - \frac{1}{2N} 
      \Tr\left(\Gamma + \Gamma^\dagger\right), \label{eq:Lambda}\\
  \Gamma &=& 
  \Tr(\Sigma^\prime B_1 U) \; Q         
+ \Tr(\Sigma^\prime B_2 U) \; Q^2 \nonumber\\
&&+ f_1  \; U \Sigma^\prime   
+ f_2  \; Q U \Sigma^\prime 
+ f_2  \; U \Sigma^\prime Q.
\label{eqn:gamma-def}
\end{eqnarray}
One sees that $\Gamma^{(k-1)}_\mu(x)$,
and hence, $\Lambda_\mu^{(k\!-\!1)}(x)$, are defined on each
lattice link in terms of the link variables $U^{(k-1)}_\mu(x)$ and 
$\Sigma^{(k)}_\mu(x)$. At this point, the details of the staple
construction in $C_\mu(x)$ must be included.  Given Eq.~(\ref{eq:Cdef})
and utilizing trace cyclicity and translational invariance,
one eventually obtains the following recursion relation:
\begin{eqnarray}
&&\Sigma_\mu(x) = \Sigma_\mu^\prime(x)
  \Bigl(f_0I+f_1Q_\mu(x)+f_2Q_\mu^2(x)\Bigr)\nonumber\\
   &&+iC^\dagger_\mu(x)\Lambda_\mu(x)  -i
 \displaystyle\sum_{\nu\neq\mu}\Bigl\{
\nonumber\\ &&
 \begin{array}[t]{l}
 \phantom{+} \rho_{\nu\mu}U_\nu(x\!+\!\hat{\mu}) 
       U^\dagger_\mu(x\!+\!\hat{\nu})U^\dagger_\nu(x)\Lambda_\nu(x) \\[1ex]
 +\rho_{\mu\nu}U^\dagger_\nu(x\!-\!\hat{\nu}\!+\!\hat{\mu})
   U^\dagger_\mu(x\!-\!\hat{\nu})\Lambda_\mu(x\!-\!\hat{\nu})
   U_\nu(x\!-\!\hat{\nu})    \\[1ex]
+\rho_{\nu\mu}U^\dagger_\nu(x\!-\!\hat{\nu}\!+\!\hat{\mu})
  \Lambda_\nu(x\!-\!\hat{\nu}\!+\!\hat{\mu})
  U^\dagger_\mu(x\!-\!\hat{\nu}) U_\nu(x\!-\!\hat{\nu}) \\[1ex]
-\rho_{\nu\mu} U^\dagger_\nu(x\!-\!\hat{\nu}\!+\!\hat{\mu})
 U^\dagger_\mu(x\!-\!\hat{\nu}) \Lambda_\nu(x\!-\!\hat{\nu})
  U_\nu(x\!-\!\hat{\nu}) \\[1ex]
-\rho_{\nu\mu}  \Lambda_\nu(x\!+\!\hat{\mu})U_\nu(x\!+\!\hat{\mu})
   U^\dagger_\mu(x\!+\!\hat{\nu})
   U^\dagger_\nu(x) \\[1ex]
+\rho_{\mu\nu}U_\nu(x\!+\!\hat{\mu})U^\dagger_\mu(x\!+\!\hat{\nu})
 \Lambda_\mu(x\!+\!\hat{\nu})
 U^\dagger_\nu(x) 
 \Bigr\},\end{array} \nonumber\\[-6mm]
\label{eq:sigma}
\end{eqnarray}
in which unprimed quantities refer to step $(k-1)$ and primed quantities
refer to step $k$.

Although the force field is related to the underlying link variables in
a complicated manner, each step in the recursive computation
of the force field outlined above is straightforward, facilitating a
natural and efficient implementation in software.  Analyticity in the
entire (finite) complex plane is an important property here; if 
$\Sigma^{(k)}_\mu(x)$ is well behaved, then $\Sigma^{(0)}_\mu(x)$ must
be also.  Note that the above formalism can be easily adapted for other
definitions of $C_\mu(x)$.

\section{Numerical tests}
\label{sec:tests}

\begin{figure}
\includegraphics[width=3.0in,bb=19 37 523 523]{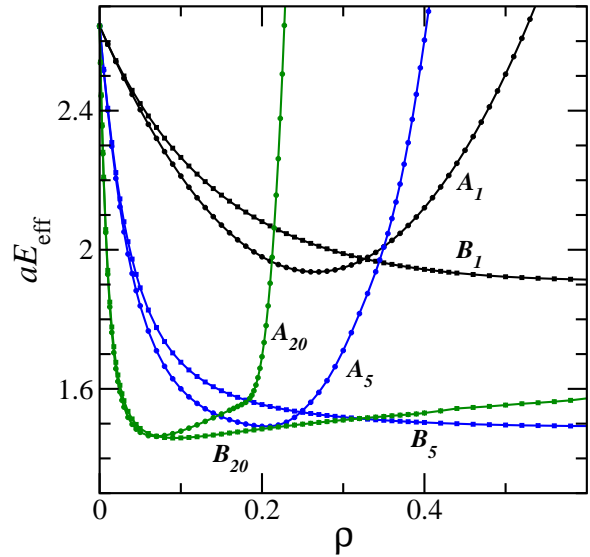}
\caption[figA]{The effective energy $aE_{\rm eff}(r)$ defined in 
 Eq.~(\ref{eq:Eeff}) for a static quark-antiquark pair separated by a 
 distance $r=5a$ for several levels of smearing $n_\rho=1,5,$ and $20$
 against the smearing parameter $\rho$.
 These results were obtained on a $12^4$ lattice using the Wilson gauge
 action with coupling $\beta=5.7$. 
 Curves labeled $A_{n_\rho}$ indicate results using the spatially-isotropic
 three-dimensional version of the analytic stout link
 smearing scheme with $n_\rho=1,5,$ and $20$ levels, while the curves
 labeled $B_{n_\rho}$ show the results for links smeared using
 Eq.~(\ref{eq:Ufuzz}) with the projection method of Ref.~\cite{su3projectA}.
\label{fig:stout_veff}}
\end{figure}

The efficacy of stout links both as a gluonic-operator smearing algorithm and
for reducing the effects of ultraviolet gluon modes in short-distance quantities
was tested in several Monte Carlo simulations.

The energy of gluons in the presence of a static quark-antiquark pair
separated by a distance $r$ can be extracted from $r\times t$ Wilson
loops $W(r,t)$ as the temporal extent $t$ becomes large.  The Wilson
loop can be viewed as the correlation function of a gauge-invariant
operator consisting of a static quark-antiquark pair connected
by a product of link variables following a straight-line path between the
quark and the antiquark.  Extraction of the lowest energy can be done much
more reliably if the couplings of the quark-antiquark-gluon operator
with higher-lying states are small.  This can be achieved by utilizing
a product of smeared links connecting the quark and the antiquark, instead
of the original link variables.  In other words, the ground state energy of
gluons in the presence of a static quark-antiquark pair can be more
reliably determined from Wilson loops $\tilde{W}(r,t)$ constructed using
smeared spatial links in the $r$-link paths on the initial and final
time slices.  A measure of how well the couplings to the higher levels are
reduced is the effective energy for a single time step, defined by
\begin{equation}
  a E_{\rm eff}(r) = -\ln\left(\frac{\tilde{W}(r,a)}{\tilde{W}(r,0)}
 \right).
 \label{eq:Eeff}
\end{equation}
Under unexceptional circumstances for a lattice gauge action with a
positive-definite transfer matrix, such as the Wilson action,
reducing the excited-state contamination tends to lower the
effective energy given above.

\begin{figure}
\includegraphics[width=3.0in,bb=34 37 523 523]{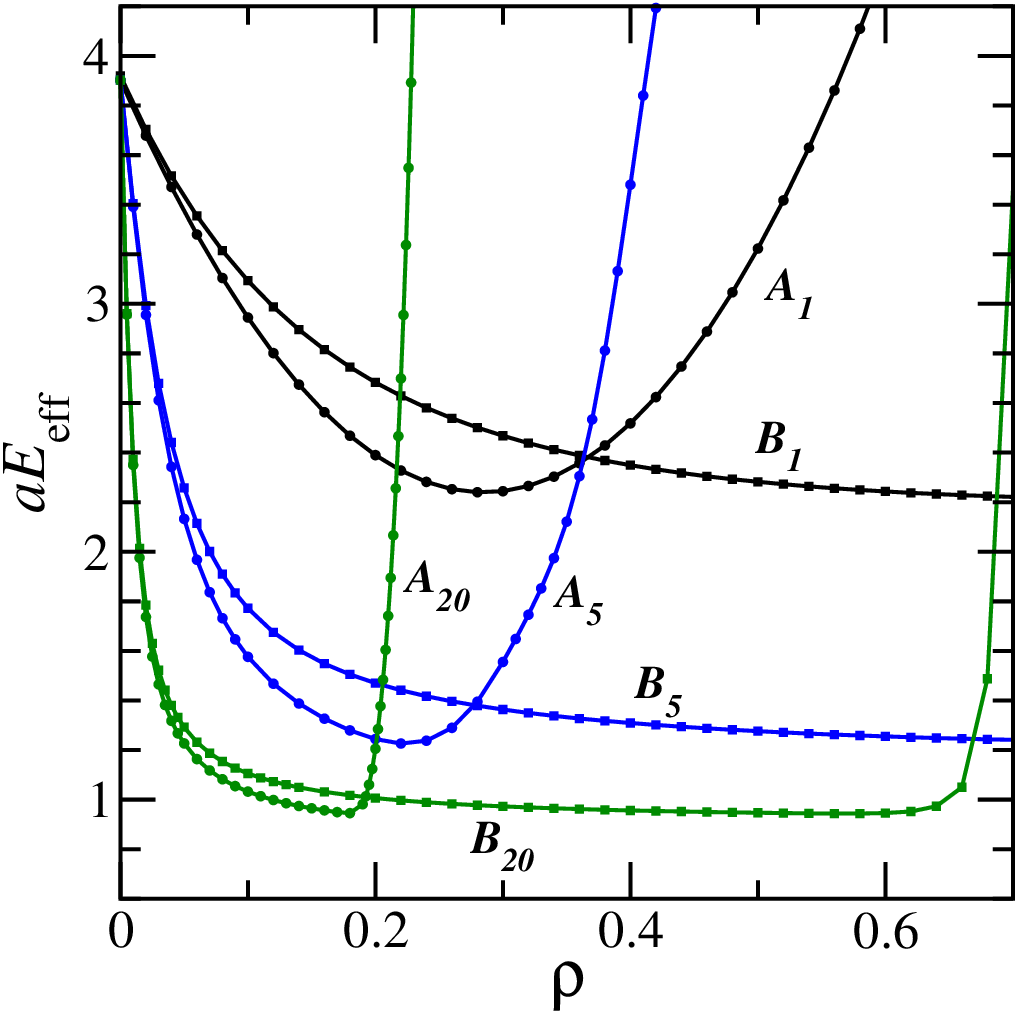}
\caption[figB]{The effective energy $aE_{\rm eff}(r)$ defined in 
 Eq.~(\ref{eq:Eeff}) for a static quark-antiquark pair separated by a 
 distance $r=10a$ for several levels of smearing $n_\rho=1,5,$ and $20$
 against the smearing parameter $\rho$.
 Results were obtained on a $24^4$ lattice using the Wilson gauge action
 with coupling $\beta=6.2$. 
 Curves labeled $A_{n_\rho}$ indicate results using the spatially-isotropic
 three-dimensional version of the analytic stout link
 smearing scheme with $n_\rho=1,5,$ and $20$ levels, while the curves
 labeled $B_{n_\rho}$ show the results for links smeared using
 Eq.~(\ref{eq:Ufuzz}) with the projection method of Ref.~\cite{su3projectA}.
\label{fig:stout_veff2}}
\end{figure}

The effective energy defined in Eq.~(\ref{eq:Eeff}) was used to test
the ability of the stout link smearing scheme to reduce excited-state
contamination in gluonic operators.   Results for $r=5a$ on a $12^4$
lattice using the Wilson gauge action with coupling $\beta=5.7$ are shown
in Fig.~\ref{fig:stout_veff}, and results for $r=10a$ on a larger $24^4$
lattice with $\beta=6.2$ are shown in Fig.~\ref{fig:stout_veff2}.
The results are compared to those obtained with links smeared using
Eq.~(\ref{eq:Ufuzz}) and the projection method of Ref.~\cite{su3projectA}.

Both figures show that link smearing can dramatically reduce
contamination from the high-lying modes of the theory.  Consider first
the results obtained using the spatially-isotropic three-dimensional
version of the stout link smearing scheme $(\rho_{jk}=\rho,
\ \rho_{4\mu}=\rho_{\mu 4}=0)$.  For a given
number of smearing iterations $n_\rho$, the effective energy decreases
initially as the smearing parameter $\rho$ is increased from zero.  
Eventually an optimal value for the smearing
parameter is reached at which the effective energy is minimized.
This optimal value decreases as the number of smearing levels $n_\rho$ is
increased, and the reduction in the effective energy at this optimal
$\rho$ value is substantial.  Further increasing $\rho$ beyond its
optimal value then results in a sharply increasing effective energy.
The rapidity of both the fall and rise of the effective energy
about its minimum is more pronounced for larger $n_\rho$. Also, the
minimum value decreases as $n_\rho$ increases until eventually a
saturation point is reached beyond which no additional reduction occurs.

The results obtained with links smeared using Eq.~(\ref{eq:Ufuzz}) and
the projection method of Ref.~\cite{su3projectA} display essentially
the same trends, except that each minimum in the effective energy is much
broader.  The increased sensitivity of the stout link smearing scheme
to the parameter $\rho$ is not surprising since $\rho$ occurs inside
an exponential function.   However, it is important to note that both
smearing methods produce nearly the same minimum values of the effective
energy.  The stout link smearing scheme is just as effective at reducing
excited-state contamination in gluonic operators as the standard
link fuzzing scheme in current use, although more careful tuning is
necessary due to the increased sensitivity to the smearing parameter $\rho$.
Note that this analytic link smearing method has already been successfully
applied in computing the spectrum of torelon excitations\cite{torelons}.

\begin{figure}
\includegraphics[width=3.0in,bb=21 37 524 523]{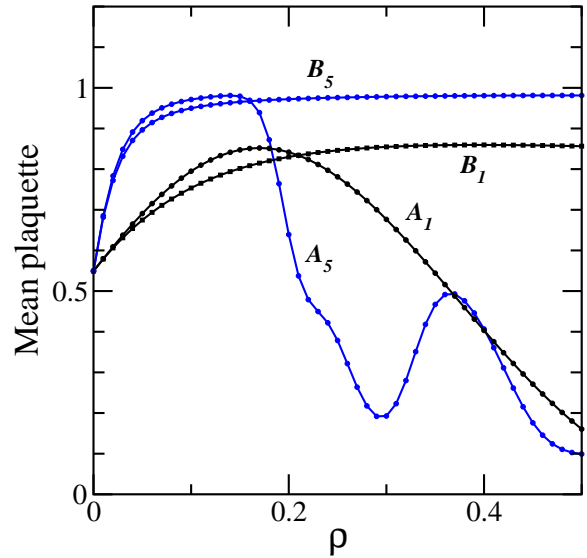}
\caption[figC]{The mean smeared plaquette against the smearing parameter
$\rho$. These results were obtained using the Wilson gauge action on a
$12^4$ lattice with coupling $\beta=5.7$. Curves labeled $A_{n_\rho}$
indicate results obtained using the isotropic four-dimensional
version of the analytic stout link
smearing scheme with $n_\rho=1$ and $5$ levels, while the curves
labeled $B_{n_\rho}$ show the results with links smeared using
Eq.~(\ref{eq:Ufuzz}) and the projection method of Ref.~\cite{su3projectA}.
\label{fig:stout_plaq}}
\end{figure}

The main purpose of using smeared links in lattice gauge and fermion
actions is to reduce discretization effects caused by the ultraviolet
modes in the lattice theory.  Often such effects are dominated by
large contributions from tadpole diagrams.  A simple measure of
how well a link smearing scheme can reduce artifacts from tadpole
diagrams is the mean smeared plaquette.  A value of the mean smeared
plaquette near unity indicates a substantial reduction of the tadpole
contributions.

Results for the mean smeared plaquette on a $12^4$ lattice using the
Wilson gauge action with coupling $\beta=5.7$ are shown in
Fig.~\ref{fig:stout_plaq}.  For isotropic four-dimensional versions
$(\rho_{\mu\nu}=\rho)$ of both smearing methods, the mean
smeared plaquette initially increases towards unity as $\rho$
increases from zero until a maximum is reached.  The two
smearing methods produce nearly the same maximum value.  As $\rho$ is
increased further, the mean smeared plaquette in the stout link scheme
quickly begins to fall, whereas the standard smearing scheme changes
little.  As $n_\rho$ initially increases from zero, the maximum value
of the mean plaquette also increases, but eventually a saturation point
is reached.  Note that the maximum values for $n_\rho=5$ are very near to
unity, suggesting a dramatic reduction of tadpole contributions. 
In summary, the analytic stout link smearing scheme is observed to be just
as efficacious as the standard smearing scheme in reducing discretization
effects from the ultraviolet gluon modes, but with an increased sensitivity
to the smearing parameter $\rho$.

\section{Conclusion}
\label{sec:conclude}

Link-variable smoothing is a crucial ingredient in constructing
gluonic operators which have dramatically reduced mixings with
the high frequency modes of the theory.  Link smearing is also
playing an increasingly important role in the construction of
improved lattice actions.  The lack of differentiability with respect
to the underlying link variables of standard smearing schemes
prevents the use of efficient Monte Carlo updating methods based
on molecular dynamics evolution.  

A link smearing method which circumvents these problems was proposed and
tested in this paper. The link smearing method is analytic everywhere in
the finite complex plane and utilizes the exponential function in such a
manner to remain within $SU(3)$, eliminating the need for any projection
back into the group.  Because of this construction, the algorithm is
also useful for any Lie group. An efficient implementation of this
smearing scheme, as well as the recursive computation of the force term
describing the change of the action in response to a variation of the link
variables, was described.  Although the force field is related to the
underlying link variables in a very complicated manner, each step
in the recursive computation of the force field is actually straightforward,
allowing a natural and efficient design in software.  The smeared mean
plaquette and the effective energy associated with the static 
quark-antiquark potential were used to show that no degradation in
effectiveness is observed as compared to link smearing
methods currently in use, although an increased sensitivity to the
smearing parameter was found.

\begin{acknowledgments}
The authors wish to thank Robert Edwards and David Richards
for helpful discussions.
This work was supported by the U.S.~National Science Foundation 
under Award PHY-0099450 and by 
Enterprise-Ireland Basic Research Grant SC/2001/306.
\end{acknowledgments}

\bibliography{cited_refs}
\end{document}